# 3D Cache Hierarchy Optimization

Leonid Yavits, Amir Morad, Ran Ginosar
Department of Electrical Engineering
Technion – Israel Institute of Technology
Haifa, Israel
yavits@tx.technion.ac.il, amirm@tx.technion.ac.il, ran@ee.technion.ac.il

*Abstract*—3D integration has the potential to improve the scalability and performance of Chip Multiprocessors (CMP). A closed form analytical solution for optimizing 3D CMP cache hierarchy is developed. It allows optimal partitioning of the cache hierarchy levels into 3D silicon layers and optimal allocation of area among cache hierarchy levels under constrained area and power budgets. The optimization framework is extended by incorporating the impact of multithreaded data sharing on the private cache miss rate. An analytical model for cache access time as a function of cache size and a number of 3D partitions is proposed and verified using CACTI simulation.

## I. Introduction

As CMP era looms, chip resources including power budget, off-chip memory bandwidth, network-on-chip (NoC) capacity and chip area remain limited [6]. Partitioning of the chip area among various CMP components, most importantly between cores and cache, but also inside the cache, to maximize CMP performance under those constraints remains a critical dilemma for the computer architect. 3D integration allows stacking DRAM above the CPU layer, alleviating the off-chip memory bandwidth constraint [6][11]. On-chip cache can also be placed in a separate silicon layer and even partitioned into a number of 3D silicon layers to improve performance. Tsai *et al.* suggested partitioning 3D cache along bit- and word-lines [17]. Loh *et al.* [12] and Puttaswamy *et al.* [16] researched bank-stacked 3D cache organization. Li *et al.* studied the performance benefits of 3D L2 cache implementation [9]. Khan *et al.* optimized 3D cache bank allocation under temperature constraint [7]. Liu *et al.* researched the benefits of CPU, cache and main memory stacking [10]. In this work we propose the optimization of 3D cache partitioning into separate 3D silicon layers under constrained area, power and NoC capacity. The framework can be easily extended to support additional constraints such as off-chip bandwidth, operating frequency, supply voltage etc.

Area allocation between cores and cache in traditional two-dimensional CMP architectures has been extensively researched. Alameldeen [1] used analytical modeling to study the trade-off between the number of cores and cache size. Oh *et al.* [15] developed detailed analytical models of various cache organizations. Krishna *et al.* [8] researched the effects of data sharing in multithreaded applications on optimal area allocation between cores and cache. In our previous study [21], we developed a framework for 2D cache hierarchy optimization. Tsai *et al.* [17] introduced 3D CACTI, an efficient tool for 3D cache access time and power estimation, including sub-array division optimization. 3D CACTI tool does not suggest however the optimal cache hierarchy and/or optimal partitioning of cache hierarchy levels into 3D silicon layers and / or area allocation among hierarchy levels. In order to find an acceptable (though potentially suboptimal) configuration, the architect has to perform a detailed design space exploration.

In our work, we aim to provide the architect with an analytical framework that yields the optimal hierarchy (the optimal number of private and shared hierarchy levels), the optimal 3D partitioning of, and the optimal area allocation among the hierarchy levels. Since average memory access delay is an additive component of the overall CMP CPI [2], [15], our results can be incorporated into a larger cores vs. cache optimization framework.

Another aspect of our work is 3D cache access time modeling. Existing studies either assume that cache access time is constant [1], [5], [15], or use CACTI [20] to simulate it. In this work we propose an analytical model for varying 3D cache access time as a function of cache size and number of its partitions. This approach leads to a more realistic cache access time model. In addition, we extend the optimization framework by modeling the effect of the multithreaded data sharing on the miss rate of the private cache.

The rest of this paper is organized as follows. Section II presents the analytical model for cache access time and its verification using 3D CACTI simulations. Section III presents the optimization framework. Section IV offers conclusions.

## II. Access Time as a Function of Cache Size and Number of Partitions

Exploring the circuit level cache model detailed in [18] and [20] while varying the Block Size, Associativity and Number of Sets, we find that the cache access time can be approximated by the analytical power-law model:

$$t_i(S_i) = \tau \cdot (\frac{S_i/\sigma}{N_{xi} \cdot N_{yi}})^\beta \qquad (1)$$

where $S_i$ is the size of the $i^{th}$ cache hierarchy level and is equal to Block Size × Associativity × Number of Sets (all variables are defined in TABLE I). In our study, we use $N_y = 1$ and vary $N_x$, based on Tsai *et al.* [17] observation that such configuration yields the best performance. The power law exponent $\beta$ is found by fitting the power law (1) curve to the cache access time data, either received by exploring circuit level models or generated by 3D CACTI. It is loosely dependent on technology node and slightly decreases with the number of 3D partitions, as shown in Fig. 1(a).

TABLE I. DEFINITIONS

| Parameter | Meaning |
|---|---|
| $\beta$ | Power law (1) exponent, range $0.4 \div 0.6$ |
| $N_{xi}, N_{yi}$ | Number of 3D partitions of $L_i$ cache hierarchy level in wordline and bitline directions, respectively |
| $\sigma, \mu, \tau, \rho$ | Baseline cache size, miss rate, access time and power consumption, respectively |
| $t_i, m_i, S_i, A_i$ [a] | $L_i$'s access time, miss rate, size and silicon area, respectively |
| $D_1, D_{12}, D_{123}$ | Average memory delays for one, two and three level cache hierarchy |
| $n$ | Number of cores in CMP |
| $A_{max}$ | Total area available to cache |
| $P_{max}$ | Total power budget available to cache |
| $M_S, M_{S\,max}$ | Rate of access and max rate of access to the shared cache, respectively |
| $d_{NoC}, d_t, d_b, d_c$ | NoC delay, comprising of transfer, blocking and queuing (congestion) delays, respectively |
| $d_D$ | DRAM access penalty |
| $E_n$ | Data sharing factor [8] |
| $\mu_N$ | Compulsory miss rate component |
| $\gamma$ | Power-law (9) exponent, range $1.35 - 1.45$ |
| $\alpha$ | Power-law (9) coefficient, $\approx 0.25$ |

[a]. Cache metrics are expressed relative to $\sigma$, $\tau$, $\mu$ and $\rho$

The power-law (1) approximates 3D CACTI simulations (and their underlying circuit level models) over a wide range of cache sizes (from 4Kbytes to 16Mbytes) to within 2%. The cache access times based on our power law model (1) vs. cache size and number of 3D layers is shown in Fig. 1(b).

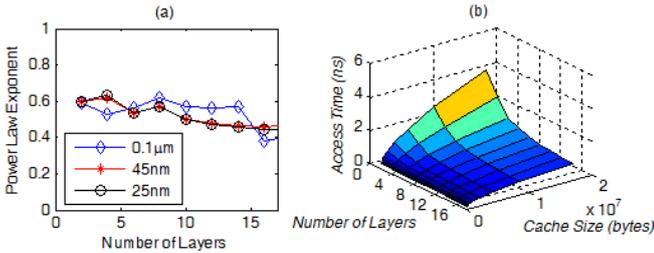

Fig. 1. (a) Power law exponent $\beta$ vs. Number of 3D Layers; (b) Access Time vs. Cache Size and Number of 3D Layers

III. OPTIMIZING CACHE HIERARCHY, 3D PARTITIONING AND AREA ALLOCATION

In this work, we focus on three typical cache configurations: single private level, two-level (one private + one shared) and three level (two private + one shared) caches. The framework can be easily extended to any number of private, shared or hybrid [15] levels. The access times of each private level $i = 1 \div 3$ and each shared level $j = 2 \div 3$ are:

$$t_i = \tau \cdot \left(S_i / \sigma N_i\right)^\beta ;$$
$$t_j = d_{NoC} + \tau \cdot \left(S_j / n\sigma N_j\right)^\beta \quad (2)$$

where $d_{NoC} = d_t + d_b + d_c$ and $N_i = N_{xi} \cdot N_{yi}$. We adopt the analytical models of NoC blocking delay $d_b$ and queuing delay $d_c$ proposed by [19] and [3]; $d_b$ and $d_c$ depend on a variety of parameters including the shared cache access rate $M_S$, the network capacity, the number of cores etc. Those parameters except for $M_S$ are not part of our optimization framework. Therefore we model both $d_b$ and $d_c$ as function of $M_S$, assuming the remaining parameters are constant. Transfer delay $d_t$ is $O(\sqrt{n})$, assuming 2-D mesh NoC [19]. The average memory delays for the above three configurations can be written as follows:

$$D_1 = (1 - m_1)t_1 + m_1(d_{NoC} + d_D);$$
$$m_1 = \mu_N + (1 - \mu_N)\mu E_n / \sqrt{S_1/\sigma} ; \quad M_S = m_1 \quad (3)$$

$$D_{12} = (1 - m_1)t_1 + m_1(1 - m_2)t_2 + m_1 m_2 d_D;$$
$$m_1 = \mu_N + (1 - \mu_N)\mu / \sqrt{S_1/\sigma} ; \quad (4)$$
$$m_2 = \mu E_n / \sqrt{(S_2 - S_1)/(n\sigma)} ; M_S = m_1;$$

$$D_{123} = (1 - m_1)t_1 + m_1(1 - m_2)t_2 + m_1 m_2(1 - m_3)t_3 + m_1 m_2 m_3 d_D;$$
$$m_1 = \mu_N + (1 - \mu_N)\mu / \sqrt{S_1/\sigma} ; \quad (5)$$
$$m_2 = \mu_N + (1 - \mu_N)\mu / \sqrt{(S_2 - S_1)/\sigma};$$
$$m_3 = \mu E_n / \sqrt{(S_3 - S_2)/(n\sigma)} ; \quad M_S = m_1 m_2$$

where $E_n$ is the data sharing factor [8], and $\mu_N$ is the compulsory miss rate component, which reflects access to data originated in remote (rather than in local) core [22]; $\mu_N$ does not depend on the size of the local cache. While $E_n$ depends on the number of cores in CMP [8], $\mu_N$ is a function of the shared data size $N$. We assume 3D DRAM implementation; hence there is no memory bandwidth restriction and no queuing delay at DRAM. Lastly, (4) and (5) assume inclusive cache and can be easily modified to support noninclusive cache.

The optimization problem can be presented as follows:

$$\text{Minimize} \quad D_o = min\{D_{123}, D_{12}, D_1\}$$
$$\text{Subject to} \quad g_j(x) \leq L_j \quad (6)$$
$$x = [S_1, S_2, S_3, N_{x1}, N_{y1}, N_{x2}, N_{y2}, N_{x3}, N_{y3}, \ldots]$$

where $D_o()$ is the objective function representing the average memory delay, yielded by the best of three possible configurations, $g_j()$ is the $j^{th}$ constrained resource, $L_j$ is the

$j^{th}$ resource limitation, and $x$ is the optimization variable vector that includes sizes of cache hierarchy levels, number of 3D partitions per cache hierarchy level and potentially additional variables. This optimization problem is solved using KKT multipliers similarly to [13]:

$$\text{Minimize} \quad D = D_o + \sum \lambda_j \cdot [g_j(x) - L_j] \quad (7)$$

where $\lambda_j$ is the KKT multiplier.

Exploring the power consumption of 3D cache using 3D CACTI, we find that it changes quite significantly depending on specific 3D configuration (for example, for $N_x \times N_y$=1×4 or 2×2 or 4×1), while remaining almost independent of the number of partitions (1, 2, 4 or 8). This result is in line with Tsai *et al.* findings (Fig. 12 in [17]). For simplicity, we assume that 3D cache power consumption does not depend on the number of 3D partitions.

It has been suggested that the power consumption of cache scales as the square root of its size [5], therefore the power constraint can be presented as function of cache size only, as follows:

$$g_1(x) = g_1(S_1, S_2, S_3) = \sum_{j=1}^{3} \rho \sqrt{S_j/\sigma} \leq P_{max} \quad (8)$$

We do not consider the power consumption of the NoC since it is common to all three configurations. Power constraint (8) is relatively basic. It does not account for the fact that a performance-optimal partitioning is not necessarily a power-optimal one [17]. In order to improve the optimization of area allocation and 3D partitioning under constrained power consumption, all possible partitions (and not only the performance-optimal ones) need to be explored; the number of subdivisions along bit- and word-lines and the number of sets per wordline ($N_{dbl}, N_{dwl}, N_{spd}$ respectively) in each silicon partition needs to be added to the optimization variable vector $x$ (6).

Exploring the design space for cache using CACTI 6.5 [14], we find that cache area can be approximated as a function of cache size by the following power law:

$$A_i = \alpha \cdot S_i^{\gamma} \quad (9)$$

where $\alpha$ and $\gamma$ are technology dependent constants calculated using MATLAB's nonlinear least square solver ***lsqcurvefit*** and presented in TABLE I. Consequently, the area constraint can be written as the sum of areas of all cache levels in all 3D silicon layers, as follows:

$$g_2(x) = g_2(S_1, S_2, S_3) = \sum_{j=1}^{3} \alpha S_j^{\gamma} \leq A_{max} \quad (10)$$

Finally, we can restrict NoC traffic by limiting the rate of access to the shared cache $M_S$:

$$g_3(x) = M_S(S_1, S_2, S_3) \leq M_{S\,max} \quad (11)$$

where $M_{S\,max}$ is the maximum capacity of the NoC.

The unconstrained objective function $D_o$ (6) can be presented in a differentiable form similarly to [13]:

$$\begin{aligned} min\{D_{123}, D_{12}, D_1\} &= min\{min\{D_{12}, D_1\}, D_{123}\} \\ &= [D_1 H_1 + D_{12}(1 - H_1)] H_2 \\ &\quad + D_{123}(1 - H_2) \end{aligned} \quad (12)$$

where $H$ is the step function:

$$H_1 = \begin{cases} 1, & D_1 < D_{12} \\ 0, & D_1 > D_{12} \end{cases}; \quad (13)$$

$$H_2 = \begin{cases} 1, & min\{D_{12}, D_1\} < D_{123} \\ 0, & min\{D_{12}, D_1\} > D_{123} \end{cases} \quad (14)$$

The partial derivatives of $H_1$ and $H_2$ with respect to $S_i$ are zero except for those $S_1, S_2$ and $S_3$ where $H_1, H_2$ and their derivatives are not defined, that is, when any of the following equalities holds:

$$D_1 = D_{12}; \; D_1 = D_{123},; \; D_{12} = D_{123} \quad (15)$$

However, these equality points are of no consequence for the optimization (since the decision there can go either way yielding the same delay), and therefore can be omitted. Consequently we can find the optimal allocation by differentiating the objective function $D$ with respect to the optimization variable vector $x$ and $\lambda_j$ and equating the resulting Jacobian vector to zero:

$$\left[ \partial D / \partial x \quad \partial D / \partial \lambda_j \right] = 0 \; \forall j = 1 \div 2 \quad (16)$$

The system of equations (16) can be solved numerically, using assumptions similar to those used in [3], [15] and [22] to find $\sigma$, $\mu$, $\rho$, $\tau$, $d_D$, $d_t$, $d_b$, $d_c$, $E_n$, $\mu_n$ and others. We apply miss rate values obtained by Krishna *et al.* [8] using PARSEC [4] and NAS [2] benchmarks.

The average memory delay vs. total cache area under constrained area, power, NoC capacity and combination of all constraints are shown in Fig. 2 (a), (b), (c) and (d) respectively, for an example of up to three levels cache and using 16 layers. The optimal area allocations per level presented as fraction of total cache area *vs.* area budget under the same constraints are shown in Fig. 3 (a), (b), (c) and (d) respectively. The optimal partitioning of cache hierarchy levels into 3D silicon layers under the same constraints is shown in Fig. 4 (a), (b), (c) and (d) respectively.

Under constrained area budget, as more area is allocated to cache, the hierarchy deepens and the average memory delay decreases to reach the optimum point (Fig. 2(a)). The entire area and the entire 3D layers stack are initially allocated to $L_1$ (Fig. 3(a), Fig. 4(a)). When area becomes sufficient for a two-level configuration, it is divided, along the 3D stack, between $L_1$ and $L_2$, with $L_1$ fraction of area and $L_1$ number of 3D layers decreasing, and $L_2$ fraction of area and number of 3D layers growing with area (Fig. 3(a), Fig. 4(a)). As area suffices for the three-level configuration, it is divided, along the 3D stack, among $L_1$, $L_2$ and $L_3$. While $L_3$ fraction of total area is continuously growing at the expense of $L_1$ and $L_2$ (Fig. 3(a)), the number of 3D silicon layers allocated to it remains low and constant (Fig. 4(a)). As area budget increases, the number of 3D layers assigned to $L_2$ grows at the expense of $L_1$ (Fig. 4(a)).

The constrained power case resembles the constrained area scenario for area below 100 (Fig. 2(b)); as area grows beyond that point, the power budget becomes insufficient for the three-level cache, so two-level hierarchy becomes optimal again (at around 110, Fig. 2(b)); as area grows further and power budget remains limited, single level cache replaces the two-level configuration as the only viable solution (at around 180, Fig. 2(b)). The area and 3D layer allocation below 100 is similar to the constrained area scenario (Fig. 3(b), Fig. 4(b)). When $L_3$ is eliminated, its area is divided between $L_1$ and $L_2$ (Fig. 3(b)), and 3D silicon layers previously occupied by it are assigned to $L_1$ (Fig. 4(b)). As area continues to grow, $L_2$ grows at the expense of $L_1$ (Fig. 3(b)), and 3D silicon layers are shifted from $L_1$ to $L_2$ (Fig. 4(b)). When $L_2$ in turn is eliminated, the entire area budget and 3D stack are allocated to $L_1$.

Under constrained NoC bandwidth, the only viable solution is the three-level configuration. No cache is possible below the minimal area required for such configuration (around 50, Fig. 2(c)). Those are the minimal area budget and the only configuration that yield $M_S$ low enough for NoC to be able to service the cache misses. As area grows, $L_1$ becomes large enough so that the single-level configuration is viable (at around 75, Fig. 2(c)). As area grows further, two-level cache becomes viable although suboptimal (at around 160, Fig. 2(c)). When the area budget is sufficient to support the three-level configuration, it is divided among $L_1$, $L_2$ and $L_3$ (around 50, Fig. 3(c)). From that point on, area allocation is similar to the constrained area scenario. Likewise, the 3D silicon layer allocation above area budget of around 50 is similar to the constrained area scenario (Fig. 4(c)).

In real life, cache performance is affected by a combination of constrained chip resources (Fig. 2(d), Fig. 3(d), Fig. 4(d)). We consider a design case where both NoC capacity and power budget are particularly limited. When cache area budget is restricted, NoC capacity is the predominant constraint, necessitating three-level cache configuration (at around 50, Fig. 2(d)), and limiting the performance of the cache. As cache area budget grows, power budget constraint becomes critical and no longer suffices to power the optimal (three-level) cache configuration (at around 110, Fig. 2(d)). When the area budget is large enough to enable the two-level configuration, the power budget is not sufficient to power it, so the single-level cache remains the only viable solution. Area and 3D silicon layer allocations resemble the constrained NoC scenario under 50 (Fig. 3(d), Fig. 4(d)). For area budget between 50 and around 110, they resemble the constrained power scenario. When the area grows above 110, the entire area and 3D silicon stack is allocated to $L_1$.

## IV. CONCLUSION

This paper describes a 3D cache hierarchy optimization framework that finds an optimal cache hierarchy and allocates hardware resources (3D silicon layers and the die area at each level) among hierarchy levels. The algorithm relies on modeling cache access time, area and power consumption as function of cache size.

The proposed framework allows performance optimization under 3D CMP restrictions such as constrained power and area budget, and NoC capacity. The optimization is extended by incorporating the impact of the multithreaded data sharing on the private cache miss rate. Our results conform to the findings of Liu *et al.* [10] that 3D memory stacking does not annul the benefits of deep cache hierarchy.

We find that under constrained area, the best performance is achieved when a smaller area budget allocated to the private (near-CPU) cache hierarchy levels is split into many 3D layers, while the larger area budget allocated to the shared hierarchy level is divided into no more than two 3D silicon layers.

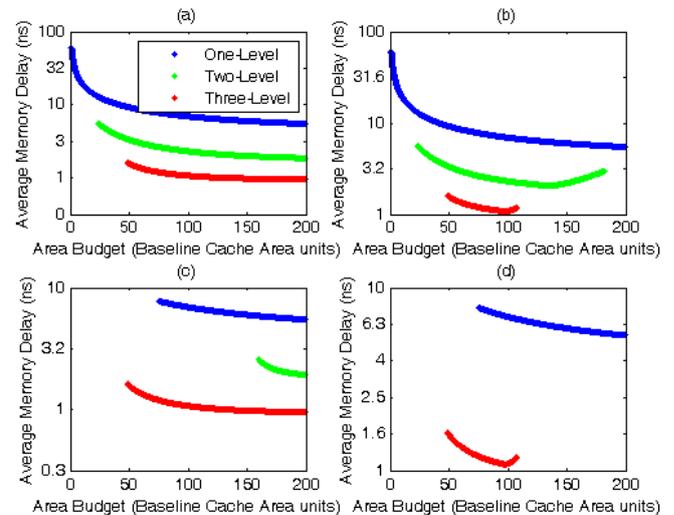

Fig. 2. Average Memory Delay *vs*. Area under: (a) constrained area, (b) constrained power, (c) constrained NoC bandwidth, (d) combined constraint.

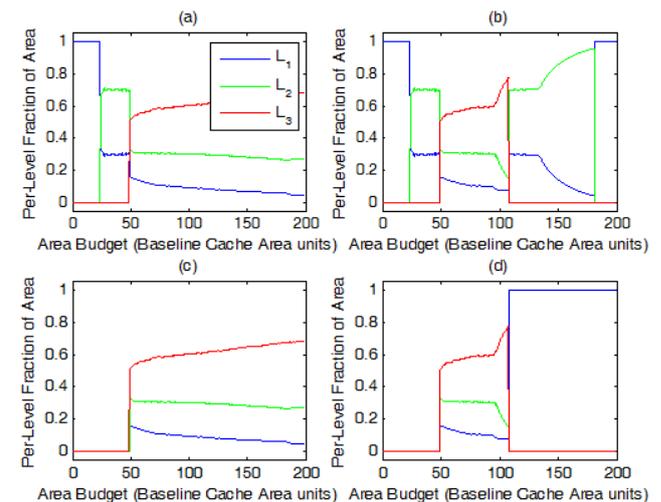

Fig. 3. Per-Cache Hierarchy Level Fraction of Area *vs*. Area under; (a) constrained area, (b) constrained power, (c) constrained NoC bandwidth, (d) combined constraint

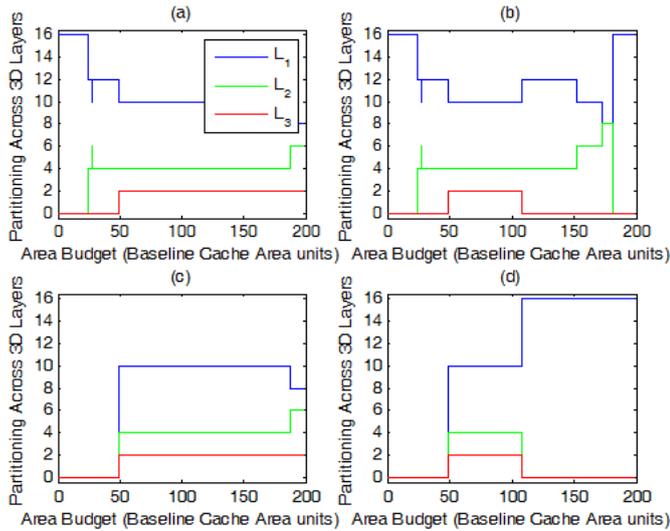

Fig. 4. Per-Cache Hierarchy Level 3D partitioning across 16 layers *vs*. Area under; (a) constrained area, (b) constrained power, (c) constrained NoC bandwidth, (d) combined constraint

We further find that in power-constrained CMPs, area overcommitment shifts the optimal configuration from three levels back to two levels, and eventually back to a single level. These results are in line with findings of Hardavellas *et al.* [6]. The 3D silicon layer and area allocation under constrained power exhibit a similar pattern, with shared levels being preferred over private ones. This happens because power consumption grows with area but is not affected by adding 3D silicon layers, while performance improves with both.

Some of our findings are counterintuitive and go against some industry conventions. For example, it is a known industry practice to make $L_1$ as large as possible as long as it can be accessed in one cycle. According to our findings, this practice may lead to suboptimal design: in NoC and area limited designs, allocating more area to $L_2$ at the expense of $L_1$ leads to better cache performance.

We have provided the architect a practical analytical tool for 3D cache hierarchy partitioning which leads to optimal memory access delay under constrained resources. This framework can be extended in a number of ways, for example it can be incorporated into the 3D CACTI simulator to offer an optimal cache hierarchy and optimal 3D resource allocation.

ACKNOWLEDGMENT

This research was partially funded by the Hasso-Plattner-Institut and the Intel Collaborative Research Institute for Computational Intelligence.